\documentclass[a4paper,12pt]{article}
\usepackage{amsmath}
\usepackage{accents}
\usepackage{bm}
\usepackage[english]{babel}
\usepackage{graphicx}
\usepackage[usenames]{color}
\usepackage{colortbl}

\begin{document}

\centerline {\LARGE{Entanglement and quantum state geometry of spin system}}
\centerline {\LARGE{with all-range Ising-type interaction}}
\medskip
\centerline {A. R. Kuzmak}
\centerline {\small \it E-Mail: andrijkuzmak@gmail.com}
\medskip
\centerline {\small \it Department for Theoretical Physics, Ivan Franko National University of Lviv,}
\medskip
\centerline {\small \it 12 Drahomanov St., Lviv, UA-79005, Ukraine}

{\small

The evolution of $N$ spin-$1/2$ system with all-range Ising-type interaction is considered. For this system we
study the entanglement of one spin with the rest spins. It is shown that the entanglement depends
on the amount of spins and the initial state. Also the geometry of manifold which contains
entangled states is obtained. For this case we find the dependence of entanglement
on the scalar curvature of manifold and examine it for different number of spins in the system. Finally
we show that the transverse magnetic field leads to a change in manifold topology.

\medskip

}

\section{Introduction\label{sec1}}

Studies of the quantum entanglement play a crucial role in the development of quantum mechanics and
quantum-information theory \cite{ENT1}. The discovery of quantum correlations allowed to solve the Einstein-Podolsky-Rosen paradox \cite{EPRP}.
In 1982, testing Bell's inequality \cite{BELL} for the entangled states of photons, EPR paradox was experimentally resolved
by Aspect et al. \cite{ASPECT}. Also Bell's inequality was tested on other systems. For instance, recently
using the electron and nuclear spins of a singled phosphorus atom in silicon a violation of Bell's inequality was demonstrated \cite{BELLINV}.
In \cite{GHZ0,GHZ} it was shown that in the case of three particles quantum correlations are much noticeable than for
two particles. The study of entanglement is closely related to the problems of preparation of entangled states. Preparation
of entangled states is required for implementation of many schemes in quantum computation. For example, the simplest scheme of quantum
teleportation of the qubit state requires the preparation of a two-qubit entangled state for the realization of a quantum channel \cite{TELEPORT}.

Nowadays the quantum entanglement of multipartite systems have been widely studied theoretically as well as experimentally \cite{EMBS}.
One of the most suitable systems for this purpose are spin systems \cite{DEODSS,ESCLLRITI,DEODIC,SCLRIQS,MEASSD,MBLIMRLRI,brach,brachass}.
The most successful realization of the multipartite entangled states was performed on the trapped ions quantum systems
\cite{SchrodCat1,EQSSTI,ITQLLWR,SchrodCat2,ETDIITIQSHI,QSDEGHTI}. An efficient  method to produce multiparticle
entangled states of ions in a trap was proposed in \cite{SchrodCat1}. This method bases on the implementation
of the effective spin system using two laser fields. The scheme for realization of such systems with the Ising and Heisenberg types
of interactions was proposed in \cite{EQSSTI}. Also in \cite{ITQLLWR} the authors suggested a scheme for the manipulation
of trapped ions using radiation in the radiofrequency or microwave regime.

To better understand the entanglement of quantum systems, the geometry property of their quantum states is often studied
\cite{LSPPTQS,GES,GESBSHF,OGES,GEMCGP,GPEBI,BGESTQ,GSESSV}. For these reasons the geometry of quantum state manifold is investigated.
This knowledge allows to obtain the information about the location of the states on quantum manifolds and simplify the problem
of their manipulation. For example, the fact that the quantum state space of spin-$1/2$ is represented by the Bloch sphere allowed to simplify
the solution of the quantum brachistochrone problem \cite{FSM2,FSM,TMTSPMF,TOSTSS}. Also information about the geometry
of quantum state manifold is important for the implementation of quantum computatations \cite{qcomp1,qcomp2,qcomp3,qcomp4}.
For example, in \cite{qcomp5} it was shown how the geometry of $n$-qubit quantum space is associated with realization of quantum computations.
In \cite{OCGQC,GAQCLB,QCAG,QGDM} it was shown that the problem of finding of the quantum circuit of unitary operators
which provide time-optimal evolution on a system of qubits is related to the problem of finding the minimal distance between two point
on the Riemannian metric.

So, in our previous papers \cite{torus,FMM}
we found the geometry of the state manifolds of two spins with different types of interaction. As a result we obtained
the conditions for the preparation of entangled states on these systems. In the present paper, we investigate the evolution (Sec. \ref{sec2})
and entanglement (Sec. \ref{sec3}) of $N$ spin-$1/2$ system with all-range Ising-type interaction. Also we calculate the metric of quantum state
manifold which contains entangled states and obtain the dependence of their entanglement on the scalar curvature of the manifold (Sec. \ref{sec4}).
The influence of the transverse magnetic field on topology of manifold is examined in Sec. \ref{sec5}.
Conclusions are presented in Sec. \ref{sec6}.

\section{The quantum evolution of spin system with all-range Ising-type interaction \label{sec2}}

We consider the $N$ spin-$1/2$ system with all-range interaction described by the Ising Hamiltonian
\begin{eqnarray}
H=\frac{J}{4}\left(\sum_{j=1}^{N}\sigma_j^z\right)^2,
\label{form1}
\end{eqnarray}
where $J$ is the interaction coupling, $N$ is the number of spins, and $\sigma_j^z$ is the Pauli matrix for spin $j$.
This Hamiltonian has $(N/2+1)$ eigenvalues in the case of even number of spins and $(N+1)/2$ in the case of odd number of spins.
These eigenvalues correspond to eigenvectors in the following way
\begin{eqnarray}
\begin{array}{ccccc}
\frac{J}{4}N^2   & \vert\uparrow\uparrow\uparrow ... \uparrow\uparrow\rangle, \vert\downarrow\downarrow\downarrow ... \downarrow\downarrow\rangle;\\[2mm]
\frac{J}{4}(N-2)^2 & \vert\downarrow\uparrow\uparrow ... \uparrow\uparrow\rangle, \vert\uparrow\downarrow\uparrow ... \uparrow\uparrow\rangle, ..., \vert\uparrow\uparrow\uparrow ... \uparrow\downarrow\rangle,\\[2mm]
         &\vert\uparrow\downarrow\downarrow ... \downarrow\downarrow\rangle, \vert\downarrow\uparrow\downarrow ... \downarrow\downarrow\rangle, ..., \vert\downarrow\downarrow\downarrow ... \downarrow\uparrow\rangle;\\[2mm]
\frac{J}{4}(N-4)^2 & \vert\downarrow\downarrow\uparrow ... \uparrow\uparrow\rangle, \vert\downarrow\uparrow\downarrow ... \uparrow\uparrow\rangle, ..., \vert\uparrow\uparrow\uparrow ... \downarrow\downarrow\rangle,\\[2mm]
         & \vert\uparrow\uparrow\downarrow ... \downarrow\downarrow\rangle, \vert\uparrow\downarrow\uparrow ... \downarrow\downarrow\rangle, ..., \vert\downarrow\downarrow\downarrow ... \uparrow\uparrow\rangle;\\[2mm]
...      & ....
\end{array}
\label{form2}
\end{eqnarray}
So, for each eigenvalue $J/4\left(N-2k\right)^2$
there exist $2\left( \begin{array}{ccccc}
N \\
k
\end{array}\right)$ eigenstates, consisting of all possible combinations of $\vert\uparrow\rangle$ and $\vert\downarrow\rangle$ states, where $k=0, \ldots, N/2$
for even $N$ and $k=0, \ldots, (N-1)/2$ for odd N.

Let us consider the evolution of the spins system having started from the initial state
\begin{eqnarray}
\vert\psi_I\rangle = \vert + + \ldots +\rangle,
\label{form5_1}
\end{eqnarray}
where
\begin{eqnarray}
\vert +\rangle = \cos\frac{\theta}{2}\vert\uparrow\rangle+\sin\frac{\theta}{2}e^{i\phi}\vert\downarrow\rangle\nonumber
\end{eqnarray}
is the state of the spin-$1/2$ projected on the positive direction of the unit vector
${\bf n}=\left(\sin\theta\cos\phi,\sin\theta\sin\phi,\cos\theta\right)$. Here $\theta$ and $\phi$ are the polar and azimuthal angles, respectively.
The action of the evolution operator with Hamiltonian (\ref{form1}) on this state is as follows
\begin{eqnarray}
\vert\psi(t)\rangle = \exp{\left(-i\frac{J}{4}\left(\sum_{j=1}^{N}\sigma_j^z\right)^2t\right)}\vert\psi_I\rangle=\sum_{k=0}^{N}c_{k}\sum_{j_1<j_2<...<j_{k}=1}^{N}\sigma_{j_1}^x\sigma_{j_2}^x...\sigma_{j_k}^x\vert\uparrow\uparrow ... \uparrow\rangle,
\label{form7_1}
\end{eqnarray}
where
\begin{eqnarray}
c_k=\cos^{N-k}\frac{\theta}{2}\sin^{k}\frac{\theta}{2}e^{ik\phi}\exp{\left(-i\frac{\chi}{4}\left(N-2k\right)^2\right)}.
\label{form8}
\end{eqnarray}
Here $\chi=Jt$. We set $\hbar=1$, which means that the energy is measured in the frequency units.
From the analysis of state (\ref{form7_1}) we obtain that it is periodic with respect to parameter $\chi\in[0,2\pi]$.

\section{The entanglement of spin system \label{sec3}}

\begin{figure}[!!h]
\centerline{\includegraphics[scale=0.5, angle=0.0, clip]{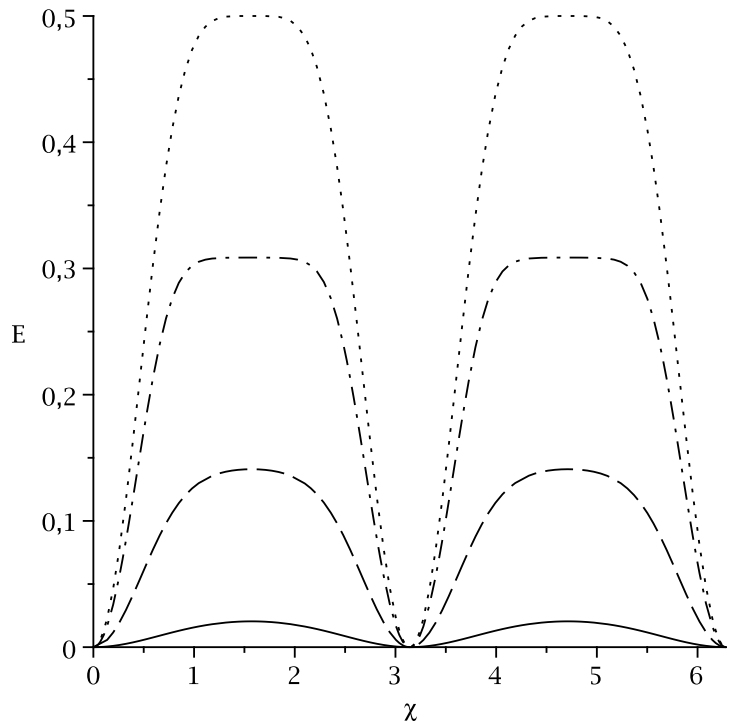}}
\caption{{\it Time dependence of entanglement of spin system having started from the initial state (\ref{form5_1}).
Results are presented for the system of six spins and different values of $\theta$: $\theta=\pi/8$ (solid curve),
$\theta=\pi/4$ (dashed curve), $\theta=3\pi/8$ (dash-dotted curve) and $\theta=\pi/2$ (dotted curve).}}
\label{ent2}
\end{figure}

In present section we calculate the entanglement of one spin with the rest spins of the system being in the state (\ref{form7_1}).
This allows us to obtain the information about the dynamics of the system through the entangled states. For this purpose we use
an expression for geometric measure of entanglement \cite{GME1,GME2,GME3,GME4} represented
by the mean value of spin \cite{EntDegree}
\begin{eqnarray}
E=\frac{1}{2}\left(1-\vert\langle{\bm \sigma}_j\rangle\vert\right),
\label{form11}
\end{eqnarray}
where in the Cartesian coordinate system $\langle{\bm \sigma}_j\rangle=\langle\sigma_j^x\rangle{\bf i}+\langle\sigma_j^y\rangle{\bf j}+\langle\sigma_j^z\rangle{\bf k}$
is the mean value of spin $j$. It is important to note that in \cite{cbeegme} the connections between geometric measure of
entanglement and relative entropy of entanglemet was studied in detail. So, calculating the mean value of spin with respect to state (\ref{form7_1})
and substituting it in (\ref{form11}) we obtain the following expression for entanglement
\begin{eqnarray}
E=\frac{1}{2}\left[1-\sqrt{\cos^2\theta+\sin^2\theta\left(\cos^2\chi+\cos^2\theta\sin^2\chi\right)^{N-1}}\right].
\label{form18}
\end{eqnarray}
From the analysis of this equation it is clear that for certain $\theta$ we achieve the maximal entangled states when
$\chi=\pi/2$ and $3\pi/2$ and the states with minimal value of entanglement ($E=0$) correspond to $\chi=0$ and $\chi=\pi$.
Therefore taking into account that $\chi\in[0,2\pi]$ we claim that for specific $\theta$ during the evolution we can reach
two maximal entangled states and two minimal entangled states (see, for example, Fig. \ref{ent2}). Also it is easy to see
that the maximal value of entanglement which can be achieved during the evolution depends on the angle $\theta$.
In the case of $\theta=0$ or $\pi$ we obtain that $E=0$. This is because the initial state is directed along the $z$-axis.
In the case of $\theta=\pi/2$ we obtain the following expression for the entanglement \cite{EDLMGM}
\begin{eqnarray}
E=\frac{1}{2}\left(1-\vert \cos\left(\chi\right)\vert^{N-1}\right).
\label{form15}
\end{eqnarray}
Then the maximal entangled state with $E=1/2$ is achieved when $\chi=\pi/2$ and $3\pi/2$.
These conditions correspond to the preparation of "Schr\"odinger cat" state \cite{SchrodCat1,SchrodCat2}.
\begin{figure}[!!h]
\centerline{\includegraphics[scale=0.5, angle=0.0, clip]{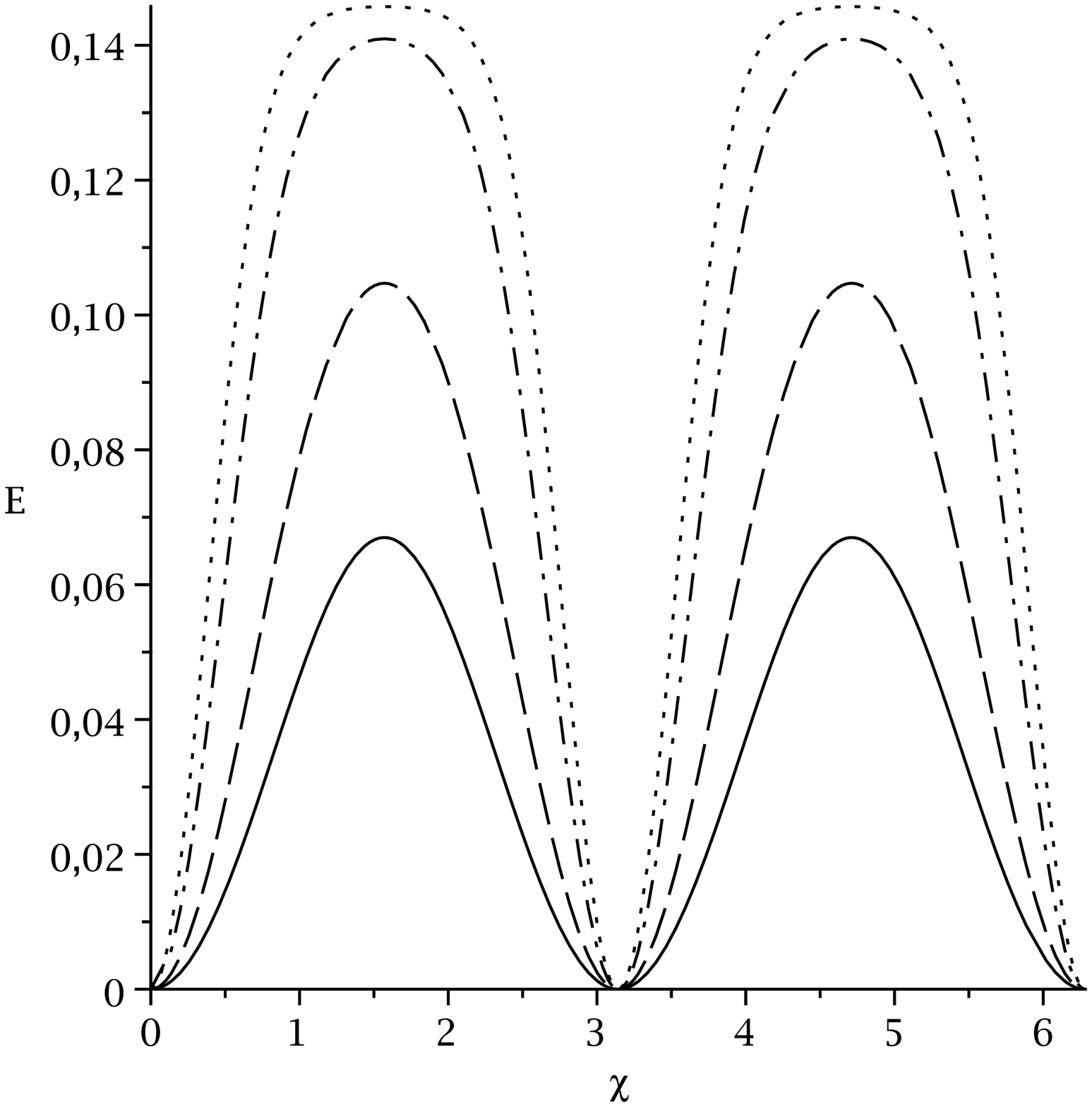}}
\caption{{\it Time dependence of entanglement (\ref{form18}) of spin system for $\theta=\pi/4$ and different numbers of spins $N$:
$N=2$ (solid curve), $N=3$ (dashed curve), $N=6$ (dash-dotted curve) and $N=9$ (dotted curve).}}
\label{ent3}
\end{figure}

It is easy to convince oneself that for an arbitrary $\theta$, excluding the cases of
$\theta=0$, $\pi/2$ and $\pi$, when the number of spins are increased, then the maximal entanglement of the system also increases.
For instance, on Fig. \ref{ent3} it is shown the properties of the system for the case of $\theta=\pi/4$.
However, for a certain $\theta$ and a large number of spins this value becomes a constant
\begin{eqnarray}
E=\frac{1}{2}\left(1-\vert\cos\theta\vert\right).
\label{form19}
\end{eqnarray}
These behaviours of the entanglement are caused by the power $N-1$ in equation (\ref{form18}). Then the entanglement is reached in time.

As we can see from (\ref{form18}), that for different $\phi$ we have the states with the same entanglement.
These states are placed on the different manifolds defined by parameters $\theta$ and $\chi$. These manifolds are geometrically similar but each of them contains
the states with a specific value of $\phi$. Let us study the geometry of these manifolds.

\section{The Fubini-Study metric of quantum state manifold \label{sec4}}

The method, which allows us to study the geometrical properties of quantum state manifold, is based on the examination of their Fubini-Study metric.
The Fubini-Study metric is defined by the infinitesimal distance $ds$ between two neighbouring pure quantum states
$\vert\psi (\xi^{\mu})\rangle$ and $\vert\psi (\xi^{\mu}+d\xi^{\mu})\rangle$ \cite{FSM2,FSM,gqev2,FSM3,FSM0,FSM1,FSMqsgm,FSMqsgm2}
\begin{eqnarray}
ds^2=g_{\mu\nu}d\xi^{\mu}d\xi^{\nu},
\label{form5}
\end{eqnarray}
where $\xi^{\mu}$ is a set of real parameters which define the state $\vert\psi(\xi^{\mu})\rangle$. The components of the metric tensor
$g_{\mu\nu}$ have the form
\begin{eqnarray}
g_{\mu\nu}=\gamma^2\Re\left(\langle\psi_{\mu}\vert\psi_{\nu}\rangle-\langle\psi_{\mu}\vert\psi\rangle\langle\psi\vert\psi_{\nu}\rangle\right),
\label{form6}
\end{eqnarray}
where $\gamma$ is an arbitrary factor which is often chosen to have the value of $1$, $\sqrt{2}$ or $2$ and
\begin{eqnarray}
\vert\psi_{\mu}\rangle=\frac{\partial}{\partial\xi^{\mu}}\vert\psi\rangle.
\label{form7}
\end{eqnarray}

Let us calculate the metric of manifold which contains state (\ref{form7_1}) with respect to $\theta$ and $\chi$.
Using definition (\ref{form5}) with expression (\ref{form6}) we obtain the following components of the metric tensor
\begin{eqnarray}
&&g_{\theta\theta}=\gamma^2\frac{N}{4},\quad g_{\chi\chi}=\frac{\gamma^2}{4}N(N-1)\sin^2\theta\left[N-1-\left(N-\frac{3}{2}\right)\sin^2\theta\right],\nonumber\\
&&g_{\theta\chi}=0.
\label{manmetric}
\end{eqnarray}
To obtain the metric tensor we calculate the following scalar products
\begin{eqnarray}
&&\langle\psi\vert\psi_{\theta}\rangle=0,\quad \langle\psi\vert\psi_{\chi}\rangle=-i\frac{N}{4}\left[1+(N-1)\cos^2\theta\right],\quad \langle\psi_{\theta}\vert\psi_{\theta}\rangle=\frac{N}{4},\nonumber\\
&&\langle\psi_{\chi}\vert\psi_{\chi}\rangle=\frac{N}{16}\left[(N-1)(N-2)(N-3)\cos^4\theta+(N-1)(6N-8)\cos^2\theta+3N-2\right],\nonumber\\
&&\langle\psi_{\theta}\vert\psi_{\chi}\rangle=i\frac{N}{4}(N-1)\sin\theta\cos\theta.
\label{scalarprod}
\end{eqnarray}

\begin{figure}[!!h]
\centerline{\includegraphics[scale=0.5, angle=0.0, clip]{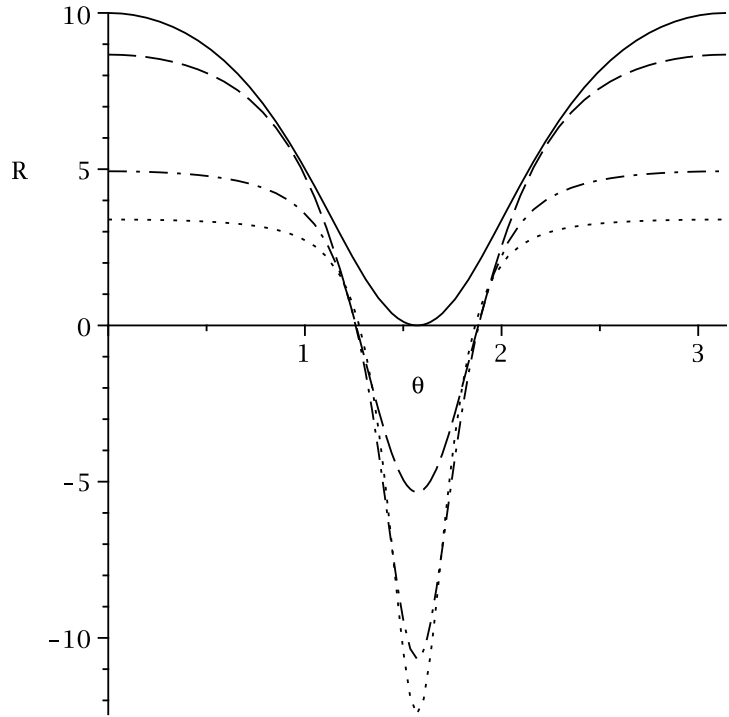}}
\caption{{\it The dependence of scalar curvature on $\theta$ (\ref{curvature2}). Results are presented for different
numbers of spins $N$:  $N=2$ (solid curve), $N=3$ (dashed curve), $N=6$ (dash-dotted curve) and $N=9$ (dotted curve). Here $\gamma=1$.}}
\label{curvature}
\end{figure}

To analyse the topology of this manifold let us calculate its scalar curvature. Using the fact that metric (\ref{manmetric})
is two-parametric and has a diagonal form, we represent the scalar curvature in the form \cite{CTF}
\begin{eqnarray}
R=\frac{2R_{\theta\chi\theta\chi}}{g_{\theta\theta}g_{\chi\chi}},
\label{curvature1}
\end{eqnarray}
where
\begin{eqnarray}
R_{\theta\chi\theta\chi}=-\frac{1}{2}\frac{\partial^2 g_{\chi\chi}}{\partial\theta^2}+\frac{1}{4g_{\chi\chi}}\left(\frac{\partial g_{\chi\chi}}{\partial\theta}\right)^2
\label{riemanntensor}
\end{eqnarray}
is the Riemann curvature tensor. Now we substitute the components of metric tensor (\ref{manmetric}) into (\ref{curvature1}) and after
simplifications obtain
\begin{eqnarray}
R=\frac{16}{\gamma^2N}\left(2-\frac{(2N-3)\cos^2\theta+N}{\left((2N-3)\cos^2\theta+1\right)^2}\right).
\label{curvature2}
\end{eqnarray}
We can see that the curvature depends on parameter $\theta$ and for $N>2$ is a negative in certain
area of manifold which satisfies the following condition
\begin{eqnarray}
\frac{(2N-3)\cos^2\theta+N}{\left((2N-3)\cos^2\theta+1\right)^2}>2.
\label{condition}
\end{eqnarray}
From the analysis of (\ref{curvature2}) we conclude that for $\theta=\pi/2$ we obtain the minimum value of curvature
\begin{eqnarray}
R_{min}=\frac{16}{\gamma^2N}(2-N)\nonumber
\end{eqnarray}
and for $\theta=0$ and $\pi$ we obtain the maximal value of curvature
\begin{eqnarray}
R_{max}=\frac{16}{\gamma^2N}\left(2-\frac{3}{4(N-1)}\right).\nonumber
\end{eqnarray}
Also the dependence of $R$ on $\theta$ is symmetric with respect to this minimum. For example, the dependence of the scalar curvature
for some $N$ is shown in Fig. \ref{curvature}

Taking into account all the above and the fact that $\theta\in[0,\pi]$ and $\chi\in[0,2\pi]$, we conclude that the manifold described by
metric (\ref{manmetric}) is closed and has a dumbbell-shape structure. It has the concave part with the centerline at $\theta=\pi/2$
and is symmetric with respect to this line. The evolution of the spin
system having started from the initial state (\ref{form5_1}) happens on the circle of radius
\begin{eqnarray}
r=\frac{\gamma}{2}\sqrt{N(N-1)}\left[N-1-\left(N-\frac{3}{2}\right)\sin^2\theta\right]^{1/2}\sin\theta.
\label{form19}
\end{eqnarray}
For certain $\theta$ (except $\theta=0$, $\pi$) this radius depends on the number of spins. The larger the number of spins, the greater the circle radius.

\begin{figure}[!!h]
\centerline{\includegraphics[scale=0.5, angle=0.0, clip]{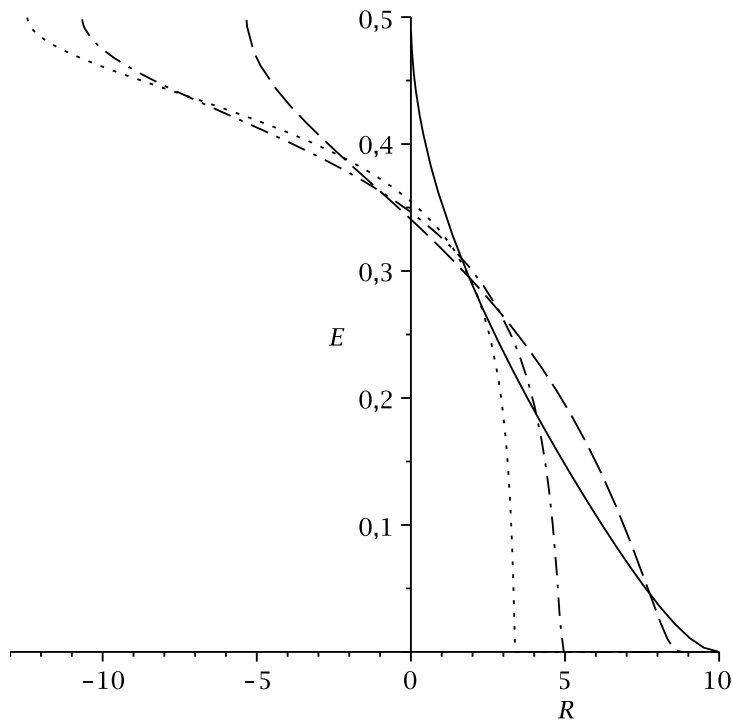}}
\caption{{\it The dependence of the entanglement on scalar curvature for $\chi=\pi/2$. Results are presented for different
numbers of spins $N$:  $N=2$ (solid curve), $N=3$ (dashed curve), $N=6$ (dash-dotted curve) and $N=9$ (dotted curve).}}
\label{ent5}
\end{figure}

We can see that entanglement (\ref{form18}) and Riemann curvature (\ref{curvature2}) depend on parameter $\theta$. This means that
the entanglement can be expressed by the Riemann curvature. For this purpose from equation (\ref{curvature2}) we find
\begin{eqnarray}
\cos^2\theta=-\frac{R\gamma^2N-24+4\sqrt{R\gamma^2N(1-N)+32N-28}}{\left(R\gamma^2N-32\right)\left(2N-3\right)}.
\label{form20}
\end{eqnarray}
So, using this expression in equation (\ref{form18}) we represent the entanglement as a function of $R$. The dependence of
entanglement on Riemann curvature for $\chi=\pi/2$ and for some number of spins is shown in Fig. \ref{ent5}.
So, the maximal value of entanglement corresponds to the minimal value of curvature and vice versa.

Using the above information we can analyse how the states with different entanglement are located on the manifold.
The manifolds with different $\phi$ are similar with respect to the distribution of the entangled states.
So, each manifold contains closed lines with respect to $\theta$. The behaviour of entanglement on this lines is defined by
equation (\ref{form18}). They have a shape of circle of radius (\ref{form19}) which for certain initial states determine
the trajectory of evolution. During the period of time $t=2\pi/J$ the state vector passes through the whole circle and returns
in the initial state.

\section{Ising model with all-range interaction in the transverse magnetic field  \label{sec5}}

In this section we study the influence of the transverse magnetic field on the geometry of quantum state manifold of the spin system
with the following Hamiltonian
\begin{eqnarray}
H=\frac{J}{4}\left(\sum_{j=1}^{N}\sigma_j^z\right)^2+\frac{h}{2}\sum_{j=1}^{N}\sigma_j^x,
\label{form21}
\end{eqnarray}
where $h$ is proportional to the value of the magnetic field. The spectrum of this model was obtained only in the thermodynamic limit
in \cite{ESLMGMTL}. Also using the semiclassical analysis and frozen-spin approximation the entanglement dynamics of this model
was explored in \cite{EDLMGM}.

So, the evolution of the system having started from state (\ref{form5_1}) can be expressed in the following form
\begin{eqnarray}
\vert\psi\rangle=\exp{\left\{-i\chi\left(\frac{1}{4}\left(\sum_{j=1}^{N}\sigma_j^z\right)^2+\frac{h}{2J}\sum_{j=1}^{N}\sigma_j^x\right)\right\}}\vert\psi_I\rangle.
\label{form22}
\end{eqnarray}
Then the Funini-Study metric takes the form
\begin{eqnarray}
&&g_{\theta\theta}=\gamma^2\frac{N}{4},\nonumber\\
&&g_{\chi\chi}=\frac{\gamma^2}{4}N\left[(N-1)\sin^2\theta\left(N-1-\left(N-\frac{3}{2}\right)\sin^2\theta\right)\right.\nonumber\\
&&\left.-2\frac{h}{J}(N-1)\sin\theta\cos^2\theta\cos\phi+\left(\frac{h}{J}\right)^2(1-\sin^2\theta\cos^2\phi)\right],\nonumber\\
&&g_{\theta\chi}=-\frac{\gamma^2}{4}\frac{h}{J}N\sin\phi.
\label{form23}
\end{eqnarray}
This metric is reduced to the diagonal form using the following transformation of the coordinates
\begin{eqnarray}
\theta=\theta'+\frac{h}{J}\sin\phi \chi',\quad\chi=\chi'.
\label{transf}
\end{eqnarray}
Then we obtain
\begin{eqnarray}
&&g_{\theta'\theta'}=\gamma^2\frac{N}{4},\nonumber\\
&&g_{\chi'\chi'}=\frac{\gamma^2}{4}N\left[(N-1)\sin^2\theta\left(N-1-\left(N-\frac{3}{2}\right)\sin^2\theta\right)\right.\nonumber\\
&&\left.-2\frac{h}{J}(N-1)\sin\theta\cos^2\theta\cos\phi+\left(\frac{h}{J}\right)^2\cos^2\theta\cos^2\phi\right].\nonumber\\
&&g_{\theta'\chi'}=0,
\label{form24}
\end{eqnarray}
where parameter $\theta$ is defined by transformation (\ref{transf}). As we can see that in this case the metric tensor depends
on the initial parameter $\phi$ and value of the magnetic field. For $h\neq 0$ and $\phi\neq\pi/2$, $-\pi/2$ the topology of
manifold is changed. In this case the metric tensor exists for any $\theta$. Then
$\theta$ takes the values in the range from $0$ to $2\pi$. The form of manifold also depends on the periodicity of evolution.
When during the some period of time the system achieves the initial state then the manifold has the form of a torus. Otherwise
we have a infinity cylinder. In the case $\phi=\pi/2$ or $-\pi/2$ the manifold has the topology of sphere.

\section{Conclusion \label{sec6}}

We considered the evolution of $N$ spin-$1/2$ system with all-range Ising-type interaction having started from the state projected
on the positive direction of the unit vector. The state which is achieved during such evolution depends on two spherical angles
which defines the initial state and the period of time of evolution. The value of entanglement of one spin with the rest spins during the
evolution was obtained. This value depends on the polar angle determined by the initial state, time of evolution and
number of spins (\ref{form18}). It was shown that for certain polar angle during the period of time $\pi/(2J)$ and $3\pi/(2J)$
the system from the disentangled initial state reaches some maximal entangled states. The value of entanglement depends on
this angle as follows: it increases when the angle changes from $0$ to $\pi/2$ and from $\pi$ to $\pi/2$. So, the maximal
possible entanglement is achieved when the initial state is located in the $xy$-plane. The achieved states are the "Schr\"odinger cat"
states \cite{SchrodCat1,SchrodCat2}.

The independence of the entanglement of the spin system on the azimuthal angle of the initial state means that there exists a continuous set
of states with the same value of entanglement. So, we studied the geometry of manifolds which contain the entangled states reached
during the evolution of the spin system. In turn, these states are placed on different manifolds defined by the polar angle
of initial states and the time of evolution. These manifolds are geometrically similar but each of them contains states with a certain
value of the azimuthal angle. We analysed the Fubini-Study metric of these manifolds and obtained that it is closed and has a dumbbell-shape
structure. Also we obtained that for a certain initial state the evolution happens along the circle of radius depending on the number of
spins (\ref{form19}). So, it can be argued that the larger the number of the spins in the system, the greater the circle radius.
We found the dependence of entanglement of the system on the scalar curvature of manifold. As a result we showed that the states are located
on the manifold so that the entanglement of the system decreases with the increasing of curvature. So, the maximally entangled states
are located on the surface with minimal curvature and vice versa.

Finnaly we studied the Fubini-Study metric of
quantum state manifold when the spin system is placed in the transverse magnetic field. We showed that the topology of manifold is changed,
except the case when the initial state is located in the perpendicular plane to the magnetic field. Then the manifold is a torus
for periodic evolution and infinity cylinder
for non-periodic evolution.

\section{Acknowledgements}

First of all, the author thanks Prof. Volodymyr Tkachuk for his great support during the investigating the problem.
The author thanks Drs. Volodymyr Pastukhov, Yuri Krynytskyi and Andrij Rovenchak for useful comments.
This work was supported in part by Project FF-30F (No. 0116U001539) from the Ministry of Education and Science of Ukraine and
by the State Fund for Fundamental Research under the project F76.


\begin{thebibliography}{99}
\bibitem{ENT1} R. Horodecki, P. Horodecki, M. Horodecki, K. Horodecki, Rev. Mod. Phys. {\bf 81}, 865 (2009).
\bibitem{EPRP} A. Einstein, B. Podolsky, N. Rosen, Phys. Rev. {\bf 47}, 777 (1935).
\bibitem{BELL} J. S. Bell, Physics {\bf 1}, 195 (1964).
\bibitem{ASPECT} A. Aspect, J. Dalibard, G. Roger, Phys. Rev. Lett. {\bf 49}, 1804 (1982).
\bibitem{BELLINV} J. P. Dehollain et al., Nature Nanotechnology {\bf 11}, 242 (2016).
\bibitem{GHZ0} D. M. Greenberg, M. A. Horne, A. Zeilinger, {\it Bell's Theorem, Quantum Theory, and Conceptions of the Universe}
(Kluwer Academics, Dordecht, The Netherlands, 1989), pp. 73-76.
\bibitem{GHZ} D. M. Greenberg, M. A. Horne, A. Shimony, A. Zeilinger, Am. J. Phys. {\bf 58}, 1131 (1990).
\bibitem{TELEPORT} C. H. Bennet, G. Brassard, C. Crepeau, R. Jozsa, A. Peres, W. K. Wootters, Phys. Rev. Lett. {\bf 70}, 1895 (1993).
\bibitem{EMBS} L. Amico, R. Fazio, A. Osterloh, V. Vedral, Rev. Mod. Phys. {\bf 80}, 517 (2008).
\bibitem{DEODSS} L. Amico, A. Osterloh, F. Plastina, R. Fazio, G. Massimo Palma, Phys. Rev. A {\bf 69}, 022304 (2004).
\bibitem{ESCLLRITI} W. D\"ur, L. Hartmann, M. Hein, M. Lewenstein, H.-J. Briegel, Phys. Rev. Lett. {\bf 94}, 097203 (2005).
\bibitem{DEODIC} G. B. Furman, V. M. Meerovich, V. L. Sokolovsky, Phys. Rev. A {\bf 77}, 062330 (2008).
\bibitem{SCLRIQS} P. Hauke, L. Tagliacozzo, Phys. Rev. Lett. {\bf 111}, 207202 (2013).
\bibitem{MEASSD} P. Neumann, N. Mizuochi, F. Rempp, P. Hemmer, H. Watanabe, S. Yamasaki, V. Jacques, T. Gaebel, F. Jelezko, J. Wrachtrup,
Science {\bf 320}, 1326 (2008).
\bibitem{MBLIMRLRI} Haoyuan Li, Jia Wang, Xia-Ji Liu, Hui Hu, Phys. Rev. A {\bf 94}, 063625 (2016).
\bibitem{brach} A. R. Kuzmak, V. M. Tkachuk, J. Phys. A \textbf{46}, 155305 (2013).
\bibitem{brachass} A. R. Kuzmak, V. M. Tkachuk, Phys. Lett. A \textbf{378}, 1469 (2014).
\bibitem{SchrodCat1} K. Molmer, A. Sorensen, Phys. Rev. Lett. {\bf 82}, 1835 (1999).
\bibitem{EQSSTI} D. Porras, J. I. Cirac, Phys. Rev. Lett. {\bf 92}, 207901 (2004).
\bibitem{ITQLLWR} F. Mintert, Ch. Wunderlich, Phys. Rev. Lett. {\bf 87}, 257904 (2001).
\bibitem{SchrodCat2} Leibried etc, Nature {\bf 438}, 639 (2005).
\bibitem{ETDIITIQSHI} J. W. Britton, B. C. Sawyer, A. C. Keith, C. C. Joseph Wang, J. K. Freericks, H. Uys, M. J. Biercuk, J. J. Bollinger,
Nature {\bf 484}, 489 (2012).
\bibitem{QSDEGHTI} J. G. Bohnet, B. C. Sawyer, J. W. Britton, M. L. Wall, A. M. Rey, M. Foss-Feig, J. J. Bollinger,
Science {\bf 352}, 1297 (2016).
\bibitem{LSPPTQS} H. A. Carterer, A. Sudbery, J. Phys. A {\bf 33}, 4981 (2000).
\bibitem{GES} M. Kus, K. \.Zyczkowski, Phys. Rev. A  {\bf 63}, 032307 (2001).
\bibitem{GESBSHF} R. Mosseri, R. Dandoloff, J. Phys. A  {\bf 34}, 10243 (2001).
\bibitem{OGES} F. Verstreate, J. Dehaene, B. De Moor, J. Mod. Opt. {\bf 49}, 1277 (2002).
\bibitem{GEMCGP} P. Levay, J. Phys. A {\bf 37}, 1821 (2004).
\bibitem{GPEBI} R. A. Bertlmann, H. Narnhofer, W. Thirring, Phys. Rev. A {\bf 66}, 032319 (2002).
\bibitem{BGESTQ} S. Ishizaka, Phys. Rev. A {\bf 69}, 020301(R) (2004).
\bibitem{GSESSV} H. Heydari, Quant. Inform. Proc. {\bf 7}, 44 (2008).
\bibitem{FSM2} I. Bengtsson and K. \.Zyczkowski, {\it Geometry of quantum states}, (New York: Cambridge University press, 2006).
\bibitem{FSM} V. M. Tkachuk, {\it Fundamental problems of quantum mechanic} (Lviv: Ivan Franko National University of Lviv, 2011). [in Ukrainian]
\bibitem{TMTSPMF} U. Boscain, P. Mason, J. Math. Phys. {\bf 47}, 062101 (2006).
\bibitem{TOSTSS} A. D. Boozer, Phys. Rev. A \textbf{85}, 012317 (2012).
\bibitem{qcomp1} A. Barenco, C. H. Bennett, R. Cleve, D. P. DiVincenzo, N. Margolus, P. Shor, T. Sleator, J. A. Smolin, H. Weinfurter, Phys. Rev. A. {\bf 52} 3457 (1995).
\bibitem{qcomp2} B. E. Kane, Nature {\bf 393}. 133 (1998).
\bibitem{qcomp3} D. P. DiVincenzo, D. Bacon, J. Kempe, G. Burkard and K. B. Whaley, Nature {\bf 408}, 339 (2000).
\bibitem{qcomp4} J. Clarke and F. K. Wilhelm, Nature {\bf 453}, 1031 (2008).
\bibitem{qcomp5} A. J. Hanson. G. Ortiz, A. Sabry and Yu-Tsung Tai, J. Phys. A {\bf 46}, 185301 (2013).
\bibitem{OCGQC} M. A. Nielsen, M. R. Dowling, M. Gu and A. C. Doherty, Phys. Rev. A {\bf 73}, 062323 (2006).
\bibitem{GAQCLB} M. A. Nielsen, Quant. Inform. Comput. {\bf 6}, 213 (2006).
\bibitem{QCAG} M. A. Nielsen, M. R. Dowling, M. Gu and A. C. Doherty, Science {\bf 311}, 1133 (2006).
\bibitem{QGDM} N. Khaneja, B. Heitmann, A. Sp\"orl, H. Yuan, T. Schulte-Herbr\"uggen and S. J. Glaser, arXiv:quant-ph/0605071 (2006).
\bibitem{torus} A. R. Kuzmak, V. M. Tkachuk, J. Phys. A \textbf{49}, 045301 (2016).
\bibitem{FMM} A. R. Kuzmak, J. Geom. Phys. \textbf{116}, 81 (2017).
\bibitem{GME1} A. Shimony, Ann. N. Y. Acad. Sci. \textbf{755}, 675 (1995).
\bibitem{GME2} D. C. Brody, L. P. Hughston, J. Geom. Phys. \textbf{38}, 19 (2001).
\bibitem{GME3} T. C. Wei, P. M. Goldbart, Phys. Rev. A. \textbf{68}, 042307 (2003).
\bibitem{GME4} L. Chen, M. Aulbach, M. Hajdusek, Phys. Rev. A \textbf{89}, 042305 (2014).
\bibitem{EntDegree} A. M. Frydryszak, M. I. Samar, V. M. Tkachuk, Eur. Phys. J. D {\bf 71}, 233 (2017).
\bibitem{cbeegme} Tzu-Chieh Wei, M. Ericsson, P. M. Goldbart, W. J. Munro, Quant. Inform. Comput. {\bf 4}, 252 (2004).
\bibitem{EDLMGM} J. Vidal, G. Palacios, C. Aslangul, Phys. Rev. A {\bf 70}, 062304 (2004).
\bibitem{gqev2} S. Abe, Phys. Rev. A {\bf 48}, 4102 (1993).
\bibitem{FSM3} S. Abe, Phys. Rev. A {\bf 46}, 1667 (1992).
\bibitem{FSM0} D. N. Page, Phys. Rev. A {\bf 36}, 3479 (1987).
\bibitem{FSM1} S. Kobayashi, K. Nomizu, {\it Fundations of Differential Geometry}, Vol. 2, Wiley, New York, (1969).
\bibitem{FSMqsgm} J. P. Provost, G. Valle, Commun. Math. Phys. {\bf 76}, 289 (1980).
\bibitem{FSMqsgm2} M. Revicule, M. Cassas, A. Plastino, Phys. Rev. {\bf 55}, 1695 (1997).
\bibitem{CTF} L. D. Landau and E. M. Lifshitz, {\it The Classical Theory of Field}, (Pergamon Press, 1971).
\bibitem{ESLMGMTL} P. Ribeiro, J. Vidal, R. Mosseri, Phys. Rev. E {\bf 78}, 021106 (2008).
\end{thebibliography}
\end{document}